\documentclass{elsart}
\usepackage{psfig}

\begin{document}
\begin{frontmatter}

  
  \title{Translationally invariant cumulants \\ in energy cascade
    models of turbulence}

\author{
Hans C.\ Eggers$^{1,2}$,
Thomas Dziekan$^{3}$
and Martin Greiner$^{2,3,4}$
}

\address{$^{1}$Department of Physics, University of Stellenbosch,
               7600 Stellenbosch, South Africa }
\address{$^{2}$Max-Planck-Institut f\"ur Physik komplexer Systeme, 
               N\"othnitzer Str.\ 38, D--01187 Dresden, Germany }
\address{$^{3}$Institut f\"ur Theoretische Physik, Technische Universit\"at,
               D--01062 Dresden, Germany }
\address{$^{4}$Department of Physics, Duke University, 
               Durham, NC 27708, USA } 

\begin{abstract}
  In the context of random multiplicative energy cascade processes, we
  derive analytical expressions for translationally invariant one- and
  two-point cumulants in logarithmic field amplitudes. Such cumulants
  make it possible to distinguish between hitherto equally successful
  cascade generator models and hence supplement lowest-order
  multifractal scaling exponents and multiplier distributions.
\end{abstract}

\end{frontmatter}

Although the underlying hydrodynamic equations are deterministic, the
statistical description of fully developed turbulence has by now a
long tradition \cite{MON71}. Random multiplicative energy cascade
models form a particularly simple and robust class of such statistical
models. While different theoretical models can reproduce
experimentally observed multifractal scaling exponents rather easily
\cite{SRE97}, observed multiplier distributions \cite{SRE95,JOU99}
eliminate many candidate cascade generators.  Nevertheless, a number
of competing generators remain, equally successful in reproducing both
scaling exponents and multipliers.  To make further progress in
ferreting out the best cascade generator within this approximation,
new observables are clearly called for.

While most experiments have concentrated on measuring statistics in
the energy dissipation density $\varepsilon$, we recently found a
complete and analytical solution working in $\ln\varepsilon$ rather
than $\varepsilon$ itself \cite{GRE98}. In this letter, we show that
cumulants in $\ln\varepsilon$ are analytically calculable even when
restoring translational invariance to the solutions to emulate the
spatial homogeneity of experimental turbulence statistics. Cumulants
turn out to be powerful tools which for third and fourth order differ
not only in magnitude but even in sign for the relevant cascade
generators and can hence be expected to distinguish between different
models which are otherwise indistinguishable in terms of observed
lowest-order multifractal scaling exponents and multiplier
distributions.

In the simplest versions of random multiplicative energy cascade
models, energy flux densities $\varepsilon$ are generated as follows:
in successive steps $j = 1,\ldots, J$, the integral scale $L$ is
divided into equal intervals of length $l_j = l_{j-1}/2 = L/2^j$ and
dyadic addresses ${\bf \kappa} = (k_1 \cdots k_j)$ with $k_i=0$ or
$1$. At each step $j$, the energy flux $\varepsilon_{k_1 \cdots k_j}$
generates fluxes in the two subintervals via
\begin{equation}
\label{rev1}
  \varepsilon_{k_1 \cdots k_j k_{j+1}}
     =  q_{k_1 \cdots k_j k_{j+1}} \;  \varepsilon_{k_1 \cdots k_j} 
        \;,
\end{equation}
where the random variables $q_L = q_{k_1 \cdots k_j 0}$ and $q_R =
q_{k_1 \cdots k_j 1}$ for the left and right subintervals are drawn
from a given cascade-generating probability density $p(q_L,q_R)$,
independently of other branches and generations of the dyadic tree.
When after $J$ cascade steps the smallest scale $\eta = l_J = L/2^J$
is reached, the local amplitudes of the flux density field
$\varepsilon_{t(\kappa)}$ are interpreted as the energy dissipation
amplitudes at positions $1 \leq t(\kappa){=}(1+\sum_{j=1}^{J} k_j
2^{J-j}) \leq 2^J$ in units of $\eta$, which are to be compared to
experimental time series converted to one-dimensional spatial series
by the frozen flow hypothesis.

We have shown previously that, since the product of multiplicative
weights $\varepsilon_{k_1 k_2 \cdots k_J} = \prod_{j=1}^J q_{k_1
  \cdots k_j}$ becomes additive on taking the logarithm,
$\ln\varepsilon_{k_1 k_2 \cdots k_J} = \sum_{j=1}^J \ln q_{k_1 \cdots
  k_j}$, the multivariate cumulant generating function for
$\ln\varepsilon$ has the analytical solution \cite{GRE98},
\begin{eqnarray}
\label{rev2}
  \ln Z(\lambda_{0 \cdots 0},\ldots,\lambda_{1 \cdots 1})
    &=&  \ln
         \left\langle  \exp\left(
         \sum_{k_1,\ldots,k_J=0}^1
         \lambda_{k_1 \cdots k_J}  \ln\varepsilon_{k_1 \cdots k_J}
         \right)  \right\rangle
         \nonumber \\
    &=&  \sum_{j=1}^J \; \sum_{k_1,\ldots,k_{j-1} = 0}^1 
         Q(\lambda_{k_1 \cdots k_{j-1}0} , \lambda_{k_1 \cdots k_{j-1}1})
         \; ,
\end{eqnarray}
where the branching cumulant generating function $Q$ (with arguments
$\lambda_{k_1 \cdots k_j} = \sum_{k_{j+1},\ldots,k_J = 0}^1
\lambda_{k_1 \cdots k_J}$) is the Mellin transform of the cascade
generator,
\begin{equation}
\label{rev3}
  Q(\lambda_L,\lambda_R)
    =  \ln\left[ \int dq_L \, dq_R \, p(q_L,q_R) \, 
       q_L^{\lambda_L} \,  q_R^{\lambda_R} \right]
       \;,
\end{equation}
which can often be found analytically. A host of analytical
predictions for statistics in $\ln\varepsilon$ for a given cascade
generator follow, starting with multivariate cumulants obtained
directly from $\ln Z$ through
\begin{equation}
\label{rev4}
  C(\kappa_1,\kappa_2,\cdots,\kappa_n)
    =  \left\langle
         \ln\varepsilon_{\kappa_1} \ln\varepsilon_{\kappa_2}
         \cdots \ln\varepsilon_{\kappa_n}
       \right\rangle_c
    =  {\partial^n \ln Z  \over
        \partial\lambda_{\kappa_1} \partial\lambda_{\kappa_2}
        \cdots \partial\lambda_{\kappa_n}
       }\biggr|_{\lambda=0} 
       \; .
\end{equation}
Due to the additivity of $\ln Z$ in (\ref{rev2}), these cumulants in
$\ln\varepsilon$ become simple sums \cite{GRE98} of same-lineage
cumulants $c_n$ and splitting cumulants $c_{r,s}$ in $\ln q$,
\begin{eqnarray}
\label{rev5}
  c_{n} 
    =  \left\langle (\ln q)^{n} \right\rangle_c
    &=&  {\partial^{n} Q  \over  \partial \lambda_L^{n}
       }\biggr|_{\lambda_L=\lambda_R=0}
       \; , \\
\label{rev6}
  c_{r,s} 
    =  \left\langle (\ln q_L)^r(\ln q_R)^s \right\rangle_c
    &=&  {\partial^{r+s} Q  \over  
        \partial \lambda_L^r  \partial \lambda_R^s
       }\biggr|_{\lambda_L=\lambda_R=0}
       \; ,
\end{eqnarray}
where without loss of generality we have assumed
$Q(\lambda_L,\lambda_R)$ to be symmetric in its arguments. When all
$n$ addresses are the same, the $n$-th order theoretical one-point
cumulant is simply
\begin{equation}
\label{rev7}
C_n(\kappa) 
=  \langle (\ln \varepsilon_\kappa)^n \rangle_c 
= Jc_n \,,
\end{equation}
while the theoretical two-point cumulant of order $(r,s)$ for bins
$\kappa_1=(k_1 \cdots k_j k_{j+1} \cdots k_J)$ and $\kappa_2=(k_1
\cdots k_j k_{j+1}^\prime \cdots k_J^\prime)$ with $k_{j+1} \neq
k_{j+1}^\prime$, separated by the ultrametric distance $D=J-j$, is
given by \cite{GRE98}
\begin{equation}
\label{rev8}
  C_{r,s}(D)
    =  \left\langle
       (\ln\varepsilon_{\kappa_1})^r (\ln\varepsilon_{\kappa_2})^s
       \right\rangle_c
    =  (J-D) c_{r+s} + (1-\delta_{D,0}) c_{r,s}
       \;.
\end{equation}
Three- and higher-point cumulants take on a form very similar to the
two-point expression \cite{GRE98}.

Before theoretical cumulants can be compared to experimentally
observed ones, two complications must be dealt with. The first is that
the generating function (\ref{rev2}) and its cumulants are not
translationally invariant, in conflict with the homogeneous statistics
characterising experimental results. The second complication arises
because experimental cumulants are derived from measured moments
rather than the other way round \cite{CAR90}, requiring translational
averaging over two-point moments rather than two-point cumulants for
theory also. The proper procedure is hence to convert theoretical
cumulants (\ref{rev8}) to moments, average these to restore
translational invariance, and then convert them back to
translationally invariant cumulants for experimental comparison.

For one-point statistics, this procedure is trivial, yielding
translationally averaged one-point cumulants $\overline C_n = Jc_n$.
Although the latter cannot distinguish between a spatially ordered
cascade process and one whose field amplitudes are spatially
randomised, useful information can nevertheless be extracted from
them. To show this we introduce three factorised model cascade
generators,
\begin{equation}
\label{rev20}
  p(q_L,q_R)
    =  p(q_L) \; p(q_R)
       \;,
\end{equation}
which have proven successful in reproducing multifractal scaling
exponents,\footnote{
  We note that other popular log-stable or log-Poisson weight
  distributions are good at reproducing multifractal scaling exponents
  but fail to account for the proper scale-correlations observed in
  multiplier distributions \cite{JOU99}.
} namely a simple binomial weight distribution
\begin{equation}
\label{rev21}
  p_{\rm binomial}(q)
    =  {\alpha_2 \over {\alpha_1+\alpha_2}} 
       \delta\left( q-(1-\alpha_1) \right)
       + {\alpha_1 \over {\alpha_1+\alpha_2}}
       \delta\left( q-(1+\alpha_2) \right)
\end{equation}
with parameters $\alpha_1=0.3$ and $\alpha_2=0.65$, a lognormal
distribution
\begin{equation}
\label{rev22}
  p_{\rm lognormal}(q)
    =  \frac{1}{\sqrt{2\pi} \sigma q}
       \exp\left[ - {1 \over 2 \sigma^2}
         \left( \ln q + \frac{\sigma^2}{2} \right)^2 
       \right]
\end{equation}
with parameter $\sigma=0.42$, and a Beta distribution
\begin{equation}
\label{rev23}
  p_{\rm Beta}(q)
    =  \frac{ \Gamma(\beta_1+\beta_2) }
            { \Gamma(\beta_1)\Gamma(\beta_2) } \;
       8^{1-\beta_1-\beta_2}
       q^{\beta_1-1} (8-q)^{\beta_2-1}
\end{equation}
with parameters $\beta_1=4.88=\beta_2/7$ and $q \in [0,8]$. All quoted
parameter values were obtained from fitting to the observed multiplier
statistics, including scale correlations \cite{JOU99}.

As shown in Fig.\ 1a, all three cascade generators yield almost
identical results for the lowest-order multifractal scaling exponents
$\tau(n)=\log_2\langle{q^n}\rangle$. Since $\langle q \rangle = 1$ by
construction, $\tau(1) = 0$ for all three distributions.  For $n=2$ we
get $\tau(2) = 0.26$ for the first two distributions and $0.23$ for
the beta distribution, i.e.\ indistinguishable within the uncertainty
of the experimental intermittency exponent $\mu = 0.25 \pm 0.05$
\cite{SRE97}. Note that the $\tau(n)$'s for the binomial and the Beta
distributions remain indistinguishable even for $n \geq 3$.

While the multiplier distributions and lowest-order scaling exponents
of the three cascade generators (\ref{rev21})--(\ref{rev23}) are hence
demonstrably indistinguishable, Fig.~1b shows that, while $c_1$ and
$c_2$ are very similar for the three models, the same-lineage
cumulants exhibit clear-cut differences in third order, with
respective values $c_3=0.05$, $0.00$ and $-0.05$.  In fourth order,
the same-lineage cumulant $c_4$ also exhibits a different sign for the
binomial and Beta distribution. For the lognormal distribution, $c_4$
is of course again zero.

The relation between the exponents and cumulants is easily
demonstrated using the cumulant branching generating function
(\ref{rev3}):
\begin{equation}
\label{rev24}
  c_n
    =  \left.
       {\partial^n Q(\lambda,0) \over \partial \lambda^n}
       \right|_{\lambda=0}
    =  \left.
       {\partial^n \ln\langle{q^\lambda}\rangle \over \partial \lambda^n}
       \right|_{\lambda=0}
    =  \left. \ln 2 \;
       {\partial^n \tau(\lambda) \over \partial \lambda^n}
       \right|_{\lambda=0}
       \;,
\end{equation}
i.e.\ the cumulant $c_n$ in $\ln{q}$ is the $n$-th derivative of the
scaling exponent $\tau(\lambda)$ at $\lambda{=}0$. In principle,
therefore, the complete set of same-lineage cumulants $c_n$ with
$1{\leq}n{<}\infty$ contains the same information as the complete set
of multifractal exponents $\tau(n)$ with $0{\leq}n{<}\infty$: for the
former case an inverse Mellin transformation recovers the (factorised)
cascade generator, while the latter uses the inverse Laplace
transform.  Once, however, we truncate the two sets of observables to
the lowest orders, $1{\leq}n{\leq}4$, they sample different aspects of
the complete information. This is the reason why the three cascade
generators given above can be hard to distinguish by means of
lowest-order multifractal scaling exponents while still exhibiting
significant differences for the third- and higher-order same-lineage
cumulants.

As mentioned previously, one-point statistics are not sensitive to the
nested spatial hierarchy of the cascade. By contrast, eq.~(\ref{rev8})
shows that two-point cumulants depend on the cascade geometry through
the ultrametric distance $D$. To access such spatial information, we
now concentrate on two-point cumulants.

For two-point moments, spatial homogeneity can be emulated by creating
a theoretical time series consisting of a chain of $m\to\infty$
independent cascade fields with $L/\eta=2^J$ finest-scale bins each
\cite{GRE97}. While this scheme is simple, it has previously played a
decisive role in explaining scale-correlations for observed multiplier
distributions \cite{SRE95,JOU99} and observed Markov properties
\cite{NAE97,CLE00}. For two-point statistics with constant bin-bin
distance $d < L/\eta$, the appropriate averaging is given by
\begin{equation}
\label{rev9}
\overline \rho_{r,s}(d)
   =  \lim_{m\to\infty} \frac{1}{(m-1)2^J} \sum_{t=1}^{(m-1)2^J}
      (\ln\varepsilon_t)^r (\ln\varepsilon_{t+d})^s
\,,
\end{equation}
where $t$ and $t{+}d$, corresponding to $\kappa_1$ and $\kappa_2$ of
(\ref{rev8}), are
in base-ten notation.  Operationally, these bins are moved over the
series in bin-sized steps, successively ``seeing'' parts of adjacent
cascade configurations. For $d<L/\eta$, the $t$-average over the infinitely
long chain of the independently and identically distributed cascade
configurations can hence be replaced by a combination of an ensemble
average $\langle \rangle$ and a $t$-average restricted to two adjacent
cascade configurations, so that (\ref{rev9}) simplifies to $\overline
\rho_{r,s}(d) = 2^{-J} \sum_{t=1}^{2^J} \rho_{r,s}(t,t+d)$, where
$\rho_{r,s}(t,t{+}d) = \langle (\ln\varepsilon_{t})^r
(\ln\varepsilon_{t{+}d})^s \rangle$.  Since furthermore the two-point
correlation density factorises whenever bins $t$ and $t+d$ belong to
independent cascade field configurations, the averaged two-point
moment finally becomes
\begin{equation}
\label{rev10}
\overline \rho_{r,s}(d)
    =  \frac{1}{2^J} \left\{
         \sum_{t=1}^{2^J-d} \rho_{r,s}(t,t+d)
         + \sum_{t=2^J-d+1}^{2^J} \rho_{r}(t) \rho_{s}(t+d)
         \right\} \,.
\end{equation}
Analytic expressions for $\rho_{r,s}(t,t{+}d)$ and $\rho_r(t) =
\langle (\ln\varepsilon_t)^r \rangle$ are readily derived by inserting
the cumulants (\ref{rev8}) into the usual relations between
$n$-variate moments and cumulants \cite{CAR90} and thence into
(\ref{rev10}).

Because they are functions of the two-point cumulants, 
two-point densities with
$d \geq 1$ also depend on the ultrametric distance $D(t,t+d)$ between
bins $t$ and $t+d$, so that (\ref{rev10}) will contain sums of the
type
\begin{equation}
\label{rev11}
  G_n(J,d) 
    =  \frac{1}{2^J}
       \sum_{t=1}^{2^J - d} \left(J - D(t,t+d)\right)^n \,.
\end{equation}
In terms of a discrete probability distribution $p(D|d,J)$,
proportional to the times each value of the $D$ is taken on while $t$
runs its course, this becomes
\begin{equation}
\label{rev12}
  G_n(J,d) 
    =  \sum_{D=1}^J p(D|d,J) \, (J-D)^n  \; ,
\end{equation}
with $p(D|d,J)$ empirically found to be
\begin{equation}
\label{rev13}
  p(D|d,J) 
    =  \left\{
       \begin{array}{lrl}
         0                 \qquad &   (1 \leq D < A)  & \\[-2mm]
         1 - (d/2^A)       \qquad & (D = A)  & \\[-2mm]
         d/2^D             \qquad &   (A <  D \leq J) & \; ,
       \end{array}
       \right.
\end{equation}
where $A = \lceil \log_2 d \rceil$ is the ceiling of $\log_2 d$.
Insertion of (\ref{rev13}) into (\ref{rev12}) leads to analytical
expressions for the geometrical coefficients
\begin{eqnarray}
\label{rev14}
  G_0(J,d) 
    &=&  (1 - 2^{-J}d)
         \; ,  \\
\label{rev15}
  G_1(J,d) 
    &=&  (J-A) - 2d(2^{-A} - 2^{-J})
         \; ,  \\
\label{rev16}
  G_2(J,d) 
    &=&  (J-A)^2 - 4d(J-A)2^{-A} +6d(2^{-A} - 2^{-J}) \,,
\end{eqnarray}
which in turn yield analytical results for the averaged two-point
densities $\overline{\rho}_{r,s}(d)$. From these, spatially
homogeneous two-point cumulants are constructed via the inversion
formulae \cite{CAR90}
\begin{eqnarray}
\label{rev17}
  \overline{C}_{1,1}(d)
    &=&  \overline{\rho}_{1,1}(d)
         - \overline{\rho}_{1}^2
         \quad ,
         \nonumber \\
  \overline{C}_{2,1}(d)
    &=&  \overline{\rho}_{2,1}(d)
         - 2 \overline{\rho}_{1} \overline{\rho}_{1,1}(d)
         - \overline{\rho}_{2} \overline{\rho}_{1}
         + 2 \overline{\rho}_{1}^3
         \quad ,
         \\
  \overline{C}_{3,1}(d)
    &=&  \overline{\rho}_{3,1}(d)
         - 3 \overline{\rho}_{1} \overline{\rho}_{2,1}(d) 
         - \overline{\rho}_{3} \overline{\rho}_{1}
         - 3 \overline{\rho}_{2} \overline{\rho}_{1,1}(d)
         + 6 \overline{\rho}_{1}^2 \overline{\rho}_{1,1}(d)
         + 6 \overline{\rho}_{2} \overline{\rho}_{1}^2
         - 6 \overline{\rho}_{1}^4
         \quad,
         \nonumber \\
  \overline{C}_{2,2}(d)
    &=&  \overline{\rho}_{2,2}(d)
         - 4 \overline{\rho}_{1} \overline{\rho}_{2,1}(d)
         - 2 (\overline{\rho}_{1,1}(d))^2
         - \overline{\rho}_{2}^2
         + 8 \overline{\rho}_{1}^2 \overline{\rho}_{1,1}(d)
         + 4 \overline{\rho}_{2} \overline{\rho}_{1}^2
         - 6 \overline{\rho}_{1}^4
         \,.
         \nonumber
\end{eqnarray}
With Eqs.\ (\ref{rev10})--(\ref{rev17}) we arrive for
$\overline C_{r,1}(d)$ at
\begin{equation}
\label{rev18}
  \overline C_{r,1}(d)
    =  G_1(J,d) \, c_{r{+}1} + G_0(J,d) \, c_{r,1} \,.
\end{equation}
This turns out to be equivalent to direct translational averaging of
(\ref{rev8}), i.e.\ $2^{-J} \sum_{t=1}^{2^J} C_{r,1}(t,t+d)$. For
$s{\neq}1$, however, such direct averaging is wrong and the full
conversion from cumulant to moment to averaged moment and back to
averaged cumulant is mandatory. For $r{=}s{=}2$ we get, for example,
\begin{eqnarray}
\label{rev19}
  \overline{C}_{2,2}(d)
    &=&  G_1(J,d) \left( c_{4} + 4c_{2}c_{1,1}  \right)
       + G_0(J,d) \left( c_{2,2} + 2c_{1,1}^2 \right) 
       \nonumber \\
    &&\   - 2 \left[ G_1(J,d) c_{2} + G_0(J,d) c_{1,1} \right]^2
       + 2 G_2(J,d) c_{2}^2
       \; , 
\end{eqnarray}
where the additional terms are a consequence of the quadratic
$(\overline{\rho}_{1,1}(d))^2$ term in the expression for $\overline
C_{2,2}(d)$ in (\ref{rev17}).

Equation (\ref{rev18}) has a simple structure, consisting of two terms
each of which is the product of a geometrical prefactor and a
branching cumulant. We note that the geometrical prefactor $G_n(J,d)$
depends only on the geometric structure of the cascade but not on the
cascade generator $p(q_L,q_R)$; in particular, as $G_n$ is independent
of the cumulant order $r$, it is the same for all $\overline C_{r,1}$.
The cascade generator, on the other hand, enters only via the
branching cumulants $c_{r+1}$ and $c_{r,1}$.

For factorised cascade generators (\ref{rev20}), eq.\ (\ref{rev18})
simplifies even further to $\overline C_{r,1}(d) = G_1(J,d)\,c_{r+1}$.
The distinguishability of $c_{r+1}$'s for the three model generators
(\ref{rev21})--(\ref{rev23}) therefore implies that two-point
cumulants $\overline C_{r,1}(d)$ of the models will differ
significantly also as they amplify $c_{r+1}$ by the geometrical
prefactor.

In order to check whether our analytical results remain statistically
significant for finite data samples, we simulated a chain of $m=10^7$
cascade configurations mimicking a time series of $m$ integral
lengths, corresponding in size to a typical experimental data set.
The length of the inertial cascade range was set to $L/\eta=2^8$. One-
and two-point densities were sampled according to Eq.~(\ref{rev9}) and
then converted into two-point cumulants using relations (\ref{rev17}).
Variances on translationally averaged moments were estimated
for each $d$ by picking, with a random $t$, just one pair of energy
fluxes $(\varepsilon_t,\varepsilon_{t+d})$ per two adjacent cascade
configurations, thereby avoiding correlations between picked pairs.
Variances for $\overline C_{r,s}$ were calculated from (\ref{rev17})
using standard error propagation.

Results for $\overline C_{2,1}$ are depicted in Fig.~2, with the solid
lines representing the analytical function $G_1(J,d)\,c_3$ and
simulations yielding the shaded bands. Based on Fig.~2, it would
appear that statistically significant values for same-lineage
cumulants $c_{r+1}$ up to at least $r=2$ can likely be extracted
experimentally. As standard errors increase with increasing order,
meaningful statements on higher orders become successively more
difficult.

While one-point cumulants $\overline C_n$ should no doubt be measured,
the two-point cumulants contain significant additional information.

Firstly, one-point cumulants cannot distinguish between the ordered
chain of cascade configurations and its spatially randomised
equivalent, while two-point cumulants can do so. This is true because
for the randomised case two-point cumulants are simply zero while in
the ordered case they retain memory of the ordered cascade tree
through the geometric coefficients $G_n(J,d)$, which according to
relation (\ref{rev12}) can be understood as $n$-th order moments of
the probability distribution $p(D|d,J)$. Thus, two-point cumulants
yield information on the cascade generator via the branching cumulants
$c_{r+1}$ as well as testing the spatially nested cascade hierarchy
via the geometric coefficients $G_n(J,d)$. To date, the probing of the
treelike structure of the underlying process has not received much
attention in the literature; we only know of Refs.\ 
\cite{MEN90,BEN97}, which discuss multifractal phase-transition-like
behaviour for two-point densities, and of Ref.\ \cite{ARN98}, which
concerns itself with wavelet scale-scale correlations.

Secondly, when cascade generators $p(q_L,q_R)$ do not factorise as in
(\ref{rev20}), the splitting cumulant $c_{r,s}$ is nonzero, and so
same-lineage cumulants plus multifractal scaling exponents are
insufficient for their complete reconstruction. Indeed, more realistic
cascade generators, attempting as best possible to reproduce the
observed one-dimensional energy dissipation data, will typically exhibit
small but nonzero residual correlations between $q_L$ and $q_R$. This
expectation follows from a projection argument \cite{JOU99}
stating that the energy cascade evolves in three dimensions but is
observed only in one. Through their dependence on a nonzero $c_{r,s}$,
the $\overline C_{r,s}$ should thus provide a more complete
characterisation of the cascade generator.

To conclude: In the pursuit of finer statistical facets of the energy
dissipation field in fully developed turbulence, translationally
averaged cumulants of logarithmic field amplitudes appear to be a
promising new tool. We have demonstrated for random multiplicative
cascade processes that two-point cumulants can be written as simple
products of geometrical coefficients times cumulant moments of the
cascade generator and can thus distinguish between cascade generators
which have more or less identical lowest-order multifractal scaling
exponents and multiplier distributions. Through their dependence on
the cascade geometry, two-point cumulants are also able to test the
treelike cascade structure.

It is tempting to apply two-point cumulants of $\ln\varepsilon$
directly to the experimental energy dissipation field deduced from
hot-wire time series to learn about the best approximate random
multiplicative cascade model and then to study possible dependences on
the Reynolds number and the flow configuration. We suggest, however,
that studies of other and more elaborate models such as hierarchical
shell models \cite{BEN97}, effects of finite inertial range etc.\ be
undertaken before this new tool is applied to data. Also, practical
problems such as choosing the finest resolution scale for the analysis
and finite sample size effects should be pondered in more detail.

The authors acknowledge fruitful discussions with J\"urgen Schmiegel,
Jochen Cleve and Jahanshah Davoudi.  This work was funded in part by
the South African National Research Foundation. HCE thanks the MPIPKS
for kind hospitality and support.


\newpage
\mbox{ }\hspace*{-15mm}
\parbox[t]{65mm}{  \psfig{file=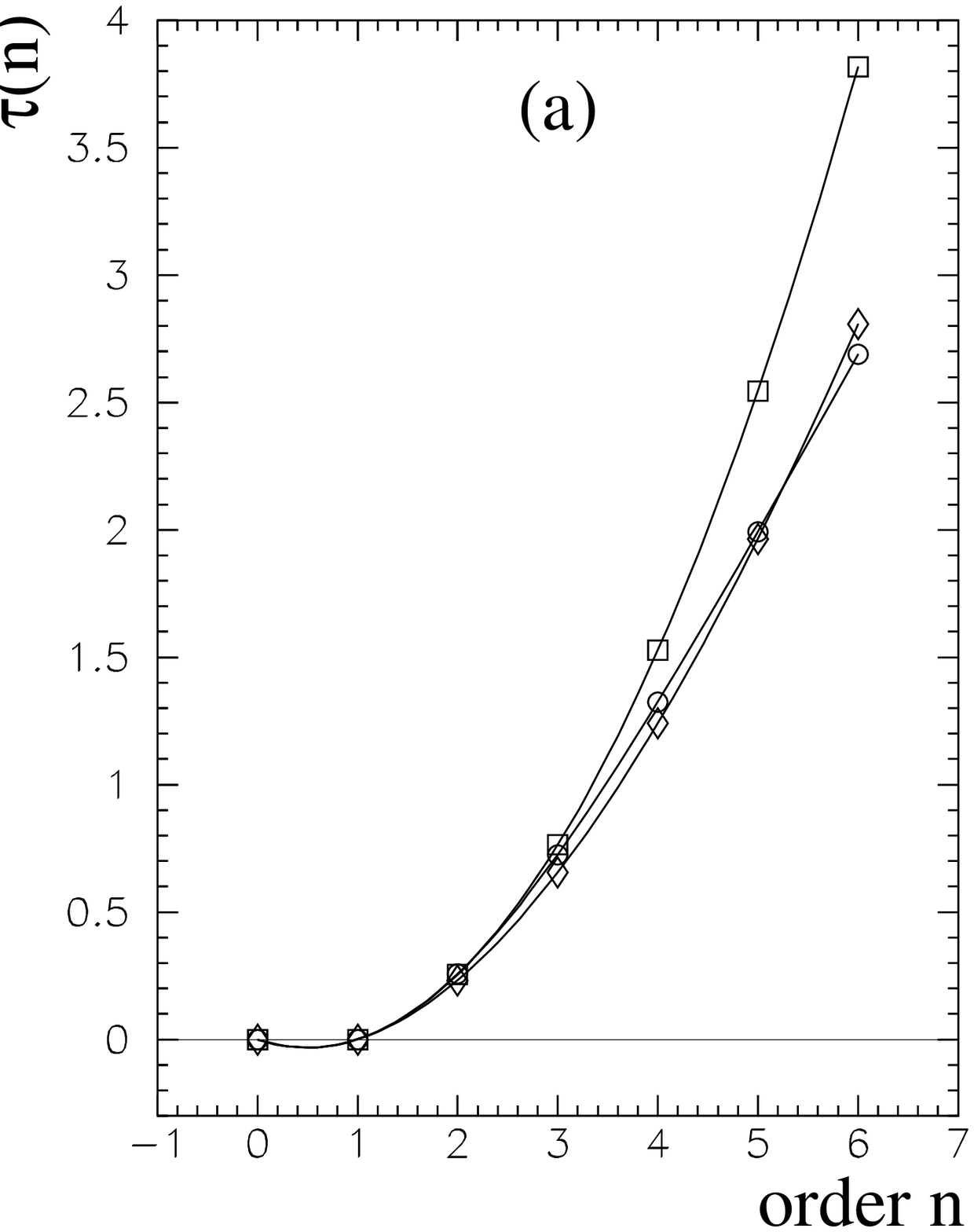,height=110mm}   }
\parbox[t]{65mm}{  \psfig{file=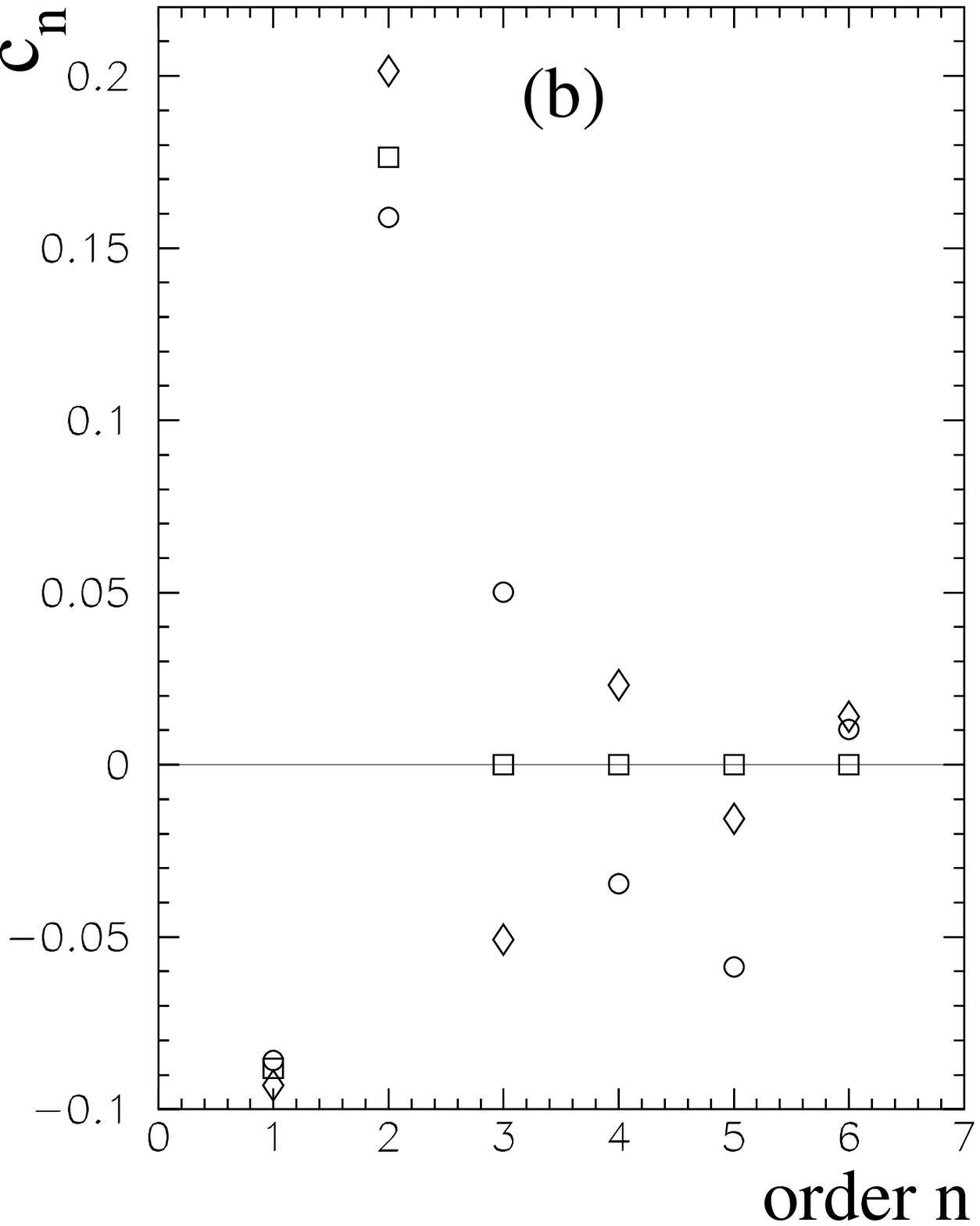,height=110mm}   }
\par\ \\
Figure 1: (a) Multifractal scaling exponents and (b) same-lineage
cumulants for the binomial (circles), lognormal (squares), and beta
(diamonds) cascade generators (\ref{rev21})--(\ref{rev23}).

\mbox{ }\hspace*{-15mm}
\parbox[t]{100mm}{  \psfig{file=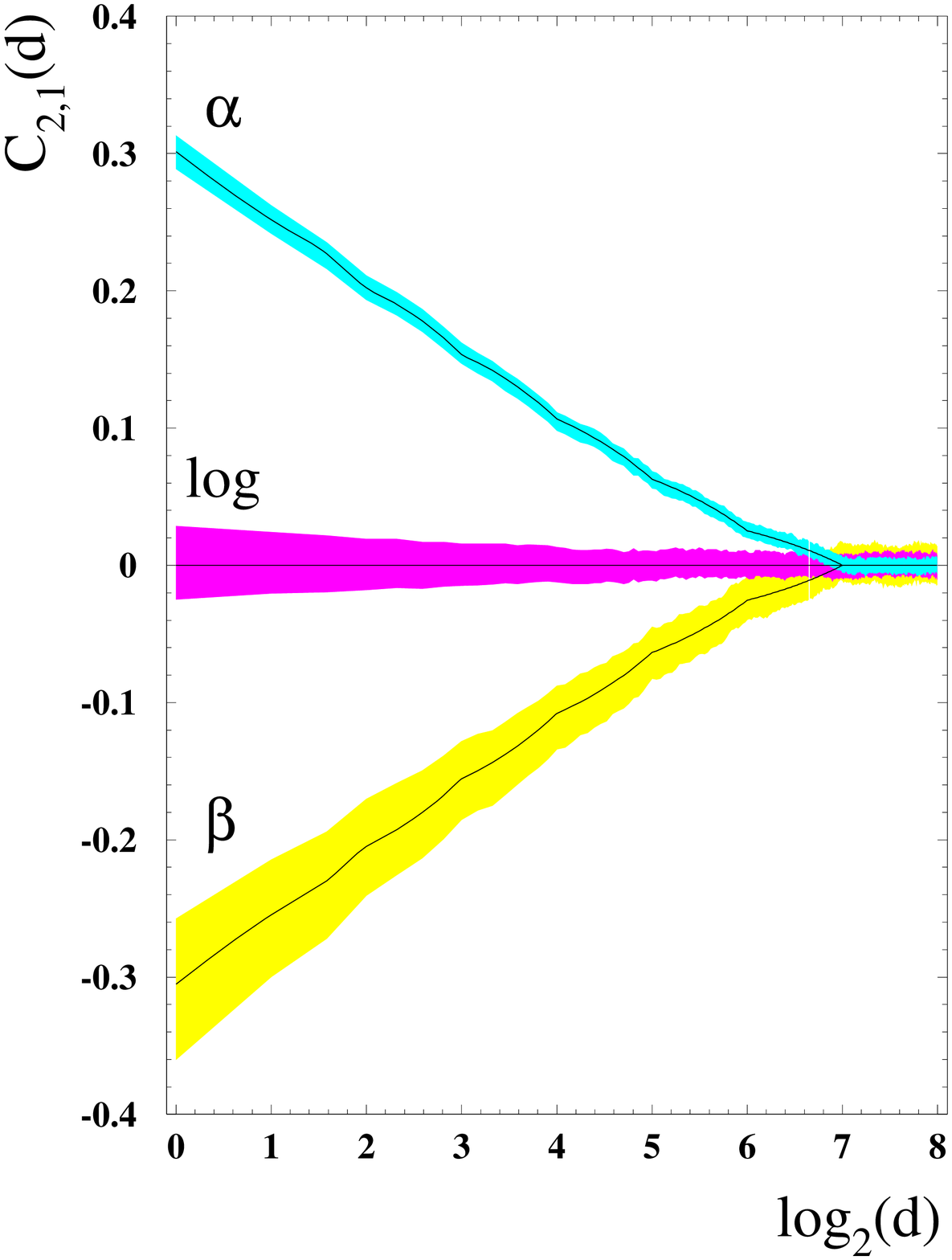,height=180mm}   }
\par\ \\
Figure 2: 
Analytical (solid lines) and simulated (shaded bands) two-point
cumulant $\overline{C}_{2,1}(d)$ as a function of bin-bin distance
$d$, for the binomial (top), lognormal (middle) and beta (bottom)
cascade generators (\ref{rev21})--(\ref{rev23}).

\end{document}